\documentclass{aa}  
\usepackage{graphicx,upgreek,txfonts,aas_macros,soul}
\usepackage{multirow,lscape}   

\usepackage{hyperref}
\usepackage{pdfpages}

\usepackage{color}
\begin{document}

\def\xmm {\emph{XMM--Newton}}
\def\cxo {\emph{Chandra}}
\def\swift {\emph{Swift}}
\def\frm {\emph{Fermi}}
\def\igr {\emph{INTEGRAL}}
\def\sax {\emph{BeppoSAX}}
\def\xte {\emph{RXTE}}
\def\rst {\emph{ROSAT}}
\def\asca {\emph{ASCA}}
\def\hst {\emph{HST}}
\def\nst {\emph{NuSTAR}}
\def\gaia {\emph{Gaia}}
\def\ngc {\mbox{NGC~6540}}
\def\src {\mbox{J1806}}
\def\flux {\mbox{erg cm$^{-2}$ s$^{-1}$}}
\def\lum {\mbox{erg s$^{-1}$}}
\def\nh {$N_{\rm H}$}
\newcommand{\rev}[1]{{#1}}
\newcommand{\revv}[1]{{#1}}
\newcommand{\revvv}[1]{{\color{blue} #1}}
\newcommand{\revvvv}[1]{{\color{blue} #1}}
\newcommand{\matt}[1]{\textcolor{red}{#1 - MI}}
\newcommand{\fabio}[1]{\textcolor{violet}{#1 - FP}}


\title{A new {\em Chandra} look at the globular cluster NGC~6540 and its peculiar X-ray flaring source}

\titlerunning{{\em Chandra} view of NGC~6540 and its flaring source}
\authorrunning{A. Sacchi et al.}

   \author{A.\,Sacchi,\inst{1,2} 
   S.\,Mereghetti,\inst{1}
   R.\,Di Stefano,\inst{2}
   J.\,A.\,Irwin,\inst{3} 
   M.\,Rigoselli,\inst{1}
   A. De Luca,\inst{1}
   N. Sims \inst{3}
    }

   \institute{
INAF--Istituto di Astrofisica Spaziale e Fisica Cosmica di Milano, via A. Corti 12, I-20133 Milano, Italy\\
e-mail: \href{mailto:andrea.sacchi@inaf.it}{andrea.sacchi@inaf.it}
\and Center for Astrophysics $\vert$ Harvard \& Smithsonian, 60 Garden Street, Cambridge, MA 02138, USA
\and Department of Physics and Astronomy, University of Alabama, Box 870324, Tuscaloosa, 35487, AL, USA
}
              
  \date{Received DD Month YYYY; accepted DD Month YYYY}

  \abstract{We report the results of a deep ($\approx65$~ks) {\it Chandra} \rev{exposure} of the globular cluster NGC~6540, obtained \revv{by} combining three observations carried out in 2023-2024  to investigate the nature of the peculiar X-ray source 3XMM J180608.9–274553. This source was previously observed with {\it XMM–Newton} to exhibit a short ($\approx300$~s) and intense X-ray flare whose luminosity and duration are inconsistent with both typical type I X-ray bursts from low-mass X-ray binaries and stellar flares. Our new data indicate three faint X-ray sources near the position of the flare seen by {\it XMM–Newton}, only one of which was detected in a previous, much shorter {\it Chandra} observation. Based on the properties of these sources -- localized at sub-arc-second precision -- and of their optical counterparts, we discuss their possible nature and association with 3XMM J180608.9–274553: only one of these newly detected sources is compatible with the position of the flare. We also discuss a few scenarios -- such as microlensing-induced amplification and black hole flaring activity analogous to that observed in Sgr~A$^\ast$ -- to explain the X-ray flare. Although these scenarios are intriguing, current observational evidence makes them both unlikely to be the origin of the \xmm\ flare.}
  \keywords{globular clusters: individual: NGC 6540, stars: flare, X-rays: binaries, X-rays: bursts}
  \maketitle
\nolinenumbers
\section{Introduction}
\label{sec:intro}

Globular clusters (GCs) are old, self-gravitationally bound stellar systems characterized by high stellar densities, particularly in their cores. Globular clusters represent excellent places to search for rare and peculiar systems of interacting compact objects, as they show an overabundance of luminous ($L_\textup{X}\gtrsim10^{36}$~erg s$^{-1}$) X-ray binaries per unit stellar mass in excess of that of the Galactic field \citep[e.g., ][]{clark75,katz75}. In the dense cores of GCs, an isolated compact object can acquire a companion through several channels: tidal capture of a main-sequence star, physical collision with a giant, or exchange encounters with one of the members of a primordial binary. Moreover, several of these systems can exhibit significant X-ray emission. The brightest configurations are low-mass X-ray binaries (LMXBs), while the faintest objects comprise a larger and more varied population of objects \citep[e.g.,~LMXBs in quiescence, millisecond pulsars, cataclysmic variables, as well as nondegenerate stars,][]{haggard09,pooley10,maxwell12,cheng18}.

Beyond these compact-object populations, GCs have also been proposed as promising hosts for intermediate-mass black holes (IMBHs; $M_{\rm BH}\sim10^2-10^5\,M_\odot$), which would occupy the elusive mass gap between stellar-mass and supermassive black holes. Their presence has been invoked to explain a handful of observational signatures, including the kinematics of stars in the innermost regions \citep[e.g., ][]{haberle24} and the tidal disruption or dynamical perturbation of compact objects hosted in the same environment \citep[e.g., ][]{tiengo22,sacchi24}. However, unambiguous detections remain elusive, and claimed candidates are frequently contested \citep[see, e.g.,][for a review]{greene20}.

\ngc\ was initially reported as an open cluster with an angular size of $\sim$0.5-1$'$ and later rediscovered as a candidate globular cluster with a possible post-collapse core by \citet{djorgovski87}. \citet{bica94} confirmed its  globular cluster nature and derived a distance of 3.5$\pm$0.4 kpc\footnote{Slightly different distances were derived by other authors: 5.2~kpc, \citet{valenti10}, 5.3~kpc \citet{harris10}, 3.02~kpc \citet{barbuy98}, and 5.91~kpc \citet{baumgardt21}. In the following, we normalize all the distance-dependent quantities to a distance $d_{{\rm 4kpc}}=(d/4~{\rm kpc})$.}.

\ngc\ hosts several X-ray sources. Among these, 3XMM~J180608.9--274553 (\src, in the following) was discovered close to the cluster center with \xmm\. 
During the EXTraS project\footnote{An EU-funded project for systematically studying the variability of all the XMM-Newton sources \citep{deluca16}.}, it was found that \src\ displayed a uniquely variable behavior \citep{mereghetti18}. In fact, on 2005 September 21, this source emitted a brief flare, lasting only $\sim$ 300 s, during which its flux increased by about two orders of magnitude, reaching a peak luminosity of $\approx10^{34}d_{{\rm 4kpc}}^2$~erg s$^{-1}$. 
This luminosity is too faint for a typical type I burst from LMXBs \citep{galloway08} and too bright for a stellar flare \citep{walter81,dempsey93}.
In this work, we present the results of a follow-up \cxo\ observation of \ngc\ and reconsider possible origins for the \src\ flare based on newly acquired data.

\section{Observations and data analysis } 
\label{sec:obs}
The first \cxo\ observation of \ngc, performed in 2008,  had a duration of only 5 ks and revealed a faint source at the position of \src\ \citep{mereghetti18}. To better study this variable source and obtain a deeper X-ray survey of \ngc, we obtained three new \cxo\ observations in 2023-2024, amounting to about 65~ks of total exposure time (see Table~\ref{tab:new_obs} for details). We employed, for all datasets, the \cxo\ Interactive Analysis of Observations (\textsc{CIAO}; v4.17.0 \citealt{fruscione2006}), using version v4.12.2 of the calibration database. We reprocessed the observations with the \texttt{chandra\_repro} task and, as all observations were performed in \texttt{VFAINT} mode, we set the \texttt{checkvfpha} flag to \texttt{yes}. 

\begin{figure}[t]
	\includegraphics[width=\hsize]{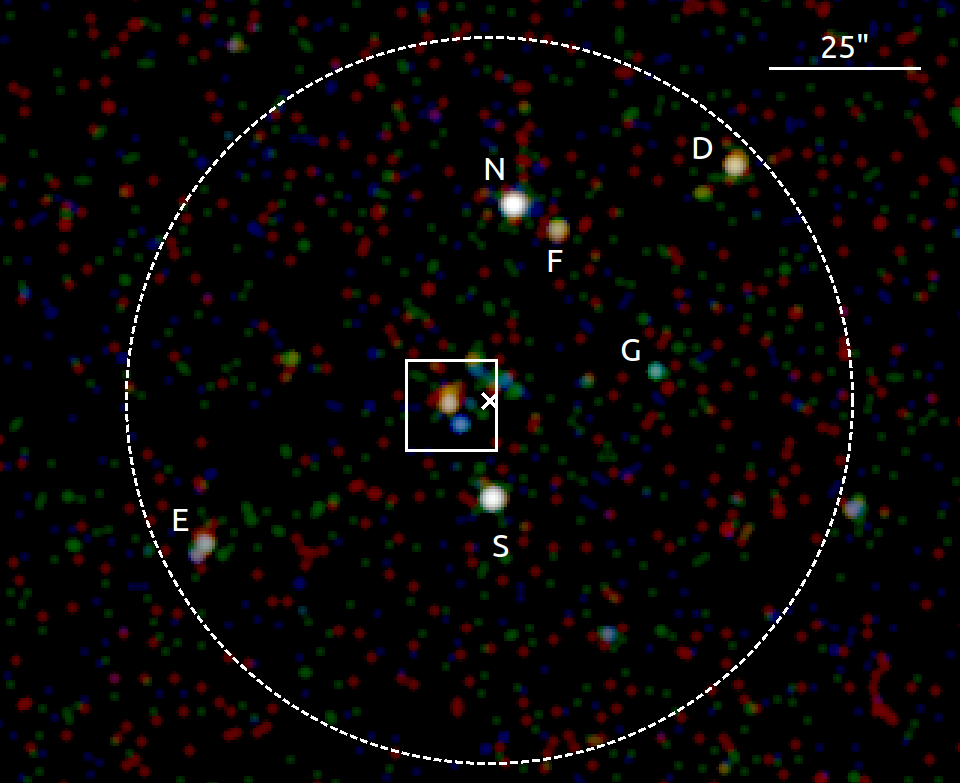}
\caption{Energy-coded \cxo\ image of \ngc , showing the soft 0.5--1.5\,keV band in red, the mid 1.5--3\,keV band, in green, and the hard 3--7\,keV band in blue. The white cross (X) and the dashed circle indicate the center and approximate extension of the cluster, respectively. The rectangle marks the region containing \src, shown in more detail in Fig.~\ref{fig:img}. North is at the top, east is to the left.
\label{fig:im_gc}}    
\end{figure}

Unless otherwise specified, for the source extractions, we employed circular regions with a radius corresponding to 90\% of the point spread function (PSF). As all the sources presented here are within 1$'$  of the instrument aim point, this amounts to $\approx2''$. The background was extracted from a circular $20''$-radius source-free region. For both timing and spectral analysis, we only considered events in the 0.5--7\,keV band. Source and background spectra, spectral redistribution matrices, and ancillary response files were generated with the routine \texttt{specextract}. 

\begin{figure}[t!]
	\includegraphics[width=\hsize]{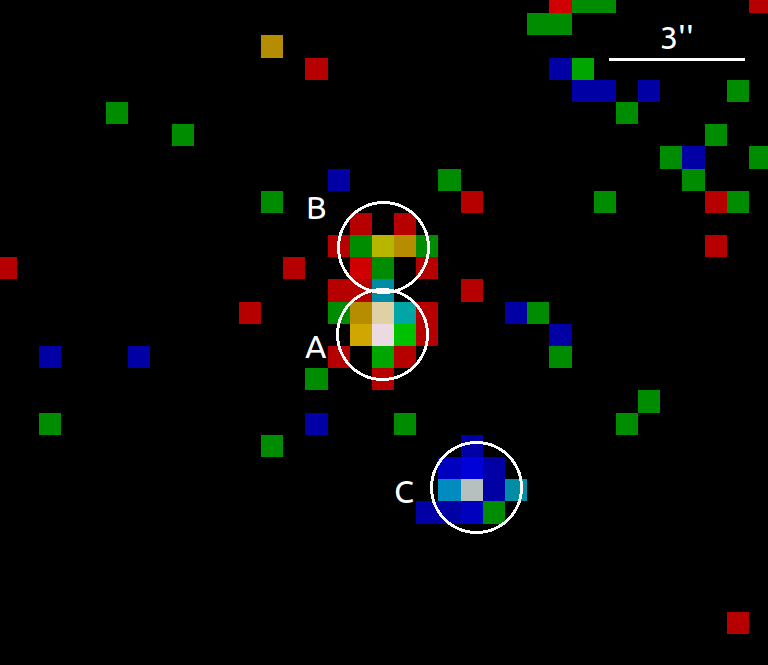}
\caption{Zoom-in of the three-color \cxo\ image on the region associated with \src. 
The circles, with a radius of $1''$, indicate the positions of the three sources within or close to the error region of \src. The color scheme is the same as in Fig.~\ref{fig:im_gc}.
\label{fig:img}}    
\end{figure}

Given that X-ray sources often exhibit significant variability, we first analyzed the three new observations individually. When statistically consistent results were provided by the individual observations, we analyzed them jointly to better constrain the properties of the detected sources.

\begin{table}
\caption{\cxo\ observation log of NGC~6540.} 
 \label{tab:new_obs}
\centering
\begin{tabular}{lccc}
\hline\hline
ObsID & Date & Exposure$^a$\\
 \hline
 8956$^b$ & 2008-11-01 & 5 \\
 27109$^c$ & 2023-03-25 & 15 \\
 27763$^c$ & 2023-03-25 & 13 \\
 26677$^c$ & 2024-08-22 & 37 \\
\hline
\end{tabular}
\tablefoot{$^a$: Approximate exposure time in kilo-second.
      $^b$: PI: Pooley.
      $^c$: PI: Irwin.} 
\end{table}

\section{Results} 
\label{sec:res}
Figure~\ref{fig:im_gc} shows a three-color composite image of \ngc\ obtained by merging the three most recent observations (total exposure of $\sim65$ ks). This was done with the \texttt{fine\_astro} and \texttt{merge\_obs} routines, employing the brightest X-ray sources in the dataset to align the three individual stares. The two routines first created exposure-corrected images and PSF maps for each dataset. They next performed a source detection (with the routine \texttt{wavdetect}) to identify all the point-like sources in each image, exploiting them to align the images. Finally, the aligned datasets were merged. As the effective area of \cxo\ has dramatically changed in the 15 years -- distinguishing this first observation from the rest of the dataset due to molecular contamination that caused a significant loss of soft sensitivity \citep{plucinsky18,plucinsky22} -- and given that the 2008 observation amounts to only 5~ks, we did not include it in our joint dataset. A zoom-in image on the central part of the cluster, where \src\ is located, is presented in Fig.~\ref{fig:img}. 

Contrary to the short ($\sim5$ ks) \cxo\ exposure from 2008, in which only a single source consistent with the location of \src\ was detected \citep{mereghetti18}, the new data reveal three distinct sources in this region. Their mutual separations on the order of $\sim1.5$–2.5$''$ are well in excess of the PSF width and of the statistical errors on their positions. Two of these sources (A and C in Fig.~\ref{fig:img}) are significantly detected in all three observations, while source B is detected only in the 2024 pointing, which has the longest exposure. As highlighted in Fig.~\ref{fig:HST}, only source~A is compatible with the position of the 2005 flare. On the contrary, the quiescent emission detected by \xmm\ is likely the blending of the emission of all three sources, considering that their fluxes and the size of \xmm's PSF are comparable. In the following subsections, we describe the properties of all three sources A, B, and C in detail.
 
Besides these three sources, six other X-ray sources were detected above a $3\sigma$-threshold by \texttt{wavdetect} in the new \cxo\ data within 1$'$ from the cluster center (see Fig.~\ref{fig:im_gc} and Table~\ref{tab:src}). In the following, we concentrate only on the three sources in the region of \src.

\subsection{Refined astrometry}
\label{sec:astro}
We corrected the astrometry of the merged dataset by cross-matching the \cxo\ point sources against a higher-precision catalog. We used the 2-micron all sky survey catalog \citep[2MASS,][the same used as reference for the Hubble Space Telescope (HST) astrometry in \citet{mereghetti18}]{skrutskies06}. To this end, we employed five \cxo\ sources. After this correction, we identified the centroid of the \cxo\ sources by fitting a 2D Gaussian profile (mimicking the PSF), with free amplitude and location, and a scale factor fixed at the PSF size at the approximate location of the source. The derived positions are given in Table~\ref{tab:src}. The errors on these positions are dominated by the uncertainty on the astrometry ($1\sigma$ error of  $0.075''$ on R.A. and  $0.099''$ on Dec), which is larger than the statistical errors derived from the Gaussian fits. 
 
\subsection{Spectral properties}
The limited photon statistics collected for the three sources in the \src\ region do not allow meaningful spectral modeling. Hence, we characterized their spectral properties using hardness ratios. We define the hardness ratio as $\mathrm{HR} = (H - S)/(H + S)$, where $S$ and $H$ are the net counts measured in the soft (0.5--2\,keV) and hard (2--7\,keV) energy bands, respectively. Given the low-count regime, uncertainties were computed using the Bayesian Estimation of Hardness Ratios code \citep[BEHR]{park06}, which properly accounts for Poisson statistics and background subtraction.

The three sources display different spectral properties. Source~C exhibits a hard spectrum with $\mathrm{HR} = 0.5^{+0.5}_{-0.5}$. This value indicates that nearly all detected counts are concentrated in the hard band. Sources~A and~B show comparatively softer emission, sharing the same hardness ratios of $\mathrm{HR} = -0.2^{+0.6}_{-0.8}$.

As spectral modeling is infeasible, as explained above, we estimated the source fluxes by converting their count rates assuming a fixed spectral model, using the online tool \texttt{PIMMS}\footnote{\url{https://cxc.harvard.edu/toolkit/pimms.jsp}}. We assumed an absorbed power-law spectrum with a fixed photon index and a column density of $N_\textup{H}=2.64\times10^{21}$~cm$^{-2}$, corresponding to the Galactic value in that direction \citep{hi4pi16}, i.e.\,, we assumed no absorption intrinsic in the GC or in the individual sources. To reflect the different HR values, the power-law photon index ($\Gamma$) was fixed at 2 for sources A and B, while a flatter value of 1.5 was assumed for source C. Additionally, to facilitate comparison with the results presented in \citet{mereghetti18}, we estimated the fluxes in the $0.5-10$~keV band employing the same power-law model they adopted to fit the quiescence of \src\ ($N_\textup{H}=6\times10^{21}$~cm$^{-2}$ and $\Gamma=2.5$). For all three sources, the fluxes inferred with the two methods have the same numerical values, as assuming a more absorbed and steeper spectrum is balanced by considering a broader energy band.

Source~A is the brightest one, with $F_{0.5-7\,\mathrm{keV}}\approx1.5\times10^{-14}$~erg~cm$^{-2}$~s$^{-1}$, corresponding to $L_X \approx3\times10^{31}~d^2_{\rm {4kpc}}$ ~erg s$^{-1}$. Source~B is the faintest one and is detected only in the August 2024 observation, with a flux $F_{0.5-7\,\mathrm{keV}}\approx8.6\times10^{-15}$~erg~cm$^{-2}$~s$^{-1}$. In previous observations, the source was undetected, and the $3\sigma$ upper limits on its flux -- obtained merging the two 2023 observations -- is $F_{0.5-7\,\mathrm{keV}}<1.6\times10^{-15}$~erg~cm$^{-2}$~s$^{-1}$, corresponding to variability by a factor of $\approx5$. Source~C has a flux of $F_{0.5-7\,\mathrm{keV}}\approx1.0\times10^{-14}$~erg~cm$^{-2}$~s$^{-1}$. 

The short 2008 \cxo\ observation detected only one source at the approximate location of source~A, with a flux compatible to the 2023-2024 level. The non-detection of sources~B and C in the short observation is consistent with the corresponding fluxes observed in 2023-2024. Given the extremely poor statistics and small separation of the three sources, we refrain from drawing any conclusion about the sources' long-term spectral variability.

\subsection{Timing properties}

As mentioned above, only source B exhibits evidence of long-term variability. No statistically significant long-term variations were seen in sources A and C when comparing their fluxes across the different observations.

We searched for variability on short timescales within the longest observation (2024); however, this analysis was also hampered by the small number of counts. To avoid arbitrary binning of the data, we tested for possible flux variations using the Kolmogorov-Smirnov (KS) test. We used it to compare the distribution of the times of arrival of each source with that expected for a constant flux. The resulting KS probabilities that the flux is constant were  0.11, 0.19, and 0.24 for sources A, B, and C, respectively.  

\begin{table*}
\caption{Properties of the X-ray sources of \ngc.} 
\label{tab:src}
\centering
\begin{tabular}{cc|cc|ccccc}
\hline\hline
Name$^a$ & Label$^b$ & R.A.$^c$ & Dec$^c$ & HR & \nh$^e$ & $\Gamma$ & Flux$^f$ & M$^g$ \\
 \hline 
                      & A & 271.53797 & --27.76534 & $-0.2^{+0.6}_{-0.8}$ & 2.6$^d$ & 2$^d$ & $1.5\pm0.2$ & 19.69\\
 2CXO J180609.1--274555 & B & 271.53798 & --27.76466 & $-0.2^{+0.6}_{-0.8}$ & 2.6$^d$& 2$^d$ & $0.9\pm0.1$ & 20.61\\
                      & C & 271.53744 & --27.76623  & $+0.5_{-0.5}^{+0.5}$ & 2.6$^d$& 1.5$^d$ & $1.0\pm0.2$ & 21.66\\
 \hline
 2CXO J180608.2--274522 & N & 271.53465 & --27.75614 & --           & $3.8\pm1.7$ & $1.3\pm0.2$ & $33.1\pm0.8$ & 21.15\\
 2CXO J180608.5--274611 & S & 271.53574 & --27.76964 & --           & $2.9\pm2.0$ & $1.2\pm0.2$ & $16.9\pm0.7$ & 20.69\\
 2CXO J180605.5--274515 & D & 271.52323 & --27.75434 & $-0.1^{+0.5}_{-0.9}$ & 2.6$^d$&  1.5$^d$ & $3.2\pm0.3$ & 17.35\\
 2CXO J180612.1--274618 & E & 271.55068 & --27.77174 & $+0.1^{+0.9}_{-0.6}$ & 2.6$^d$ & 1.5$^d$  & $1.6\pm0.2$ & 19.56\\
 2CXO J180607.7--274526 & F & 271.53235 & --27.75729 & $-0.1^{+0.6}_{-0.9}$ & 2.6$^d$ & 1.5$^d$ & $1.3\pm0.2$ & 13.57\\
  --                   & G & 271.52733 & --27.76376 & $+0.2^{+0.8}_{-0.6} $ & 2.6$^d$ & 1.5$^d$  & $0.9\pm0.1$ & 18.43\\
\hline
\end{tabular}
\tablefoot{\small Errors are at the 1$\sigma$ confidence level.
      $^a$: IAU name if available.
      $^b$: Label as reported in Fig.\ref{fig:im_gc}.
      $^c$: Astrometric corrected J2000 coordinates. The $1\sigma$ error on R.A. is $0.075"$ and on DEC is $0.099"$.
      $^d$: The parameter was fixed during the fitting procedure.
      $^e$: \nh\ column density in units of $10^{21}$~atoms/cm$^2$.
      $^f$: 0.5--7~keV unabsorbed X-ray flux in units of $10^{-14}$~erg~cm$^{-2}$~s$^{-1}$ averaged on the considered observations, except for source B, whose reported flux was obtained from the most recent observation; see text for details.
      $^g$: AB magnitude from the HST/ACS with F606W filter, except for source F, for which HST/WFC3 F160W AB magnitude is listed. The magnitudes were retrieved from the Hubble Legacy Archive (HLA; \url{https://hla.stsci.edu/hlaview.html}).} 
\end{table*}

\begin{figure}
	\includegraphics[width=\hsize]{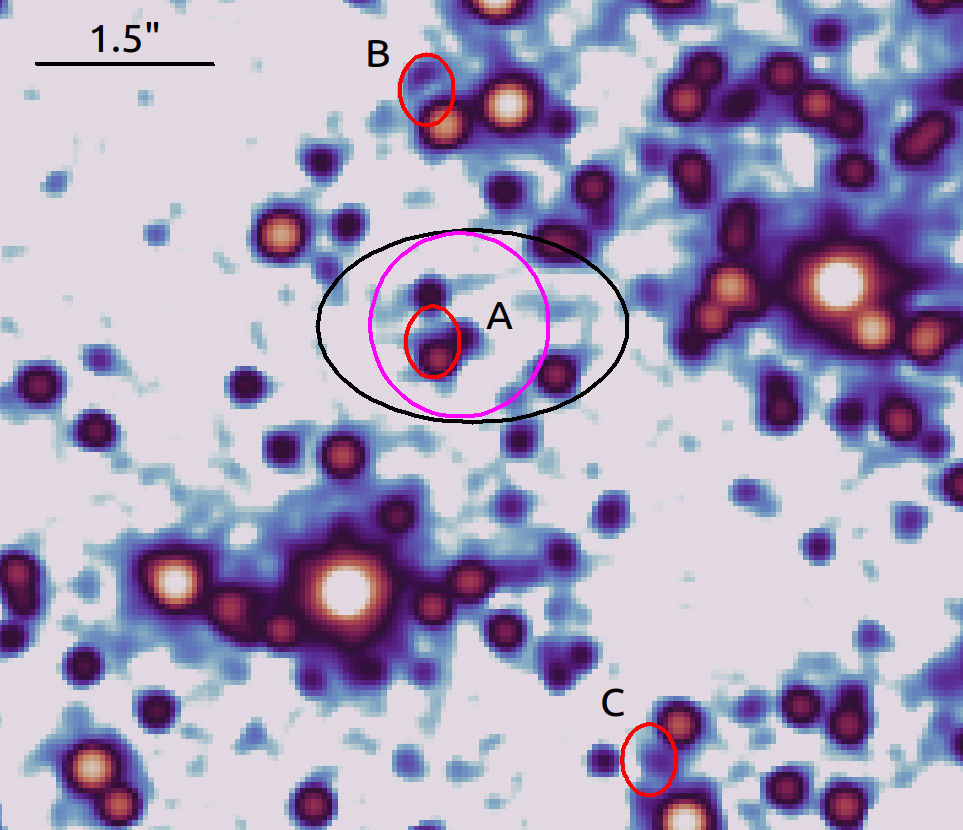}
\caption{HST (F555W) image of the central region of \ngc . The red ellipses represent the positions derived in this work for sources A, B, and C (3$\sigma$ errors). The black ellipse represents the 99\% uncertainty region for source A obtained with the short \cxo\ observation from 2008 \citep{mereghetti18}.  The magenta circle represents the 1$\sigma$ uncertainty region of \src\ derived from our reanalysis of the \xmm\ data, as described in Sect. \ref{xmm_pos}.  
\label{fig:HST}}    
\end{figure}

\begin{figure}
	\includegraphics[width=\hsize]{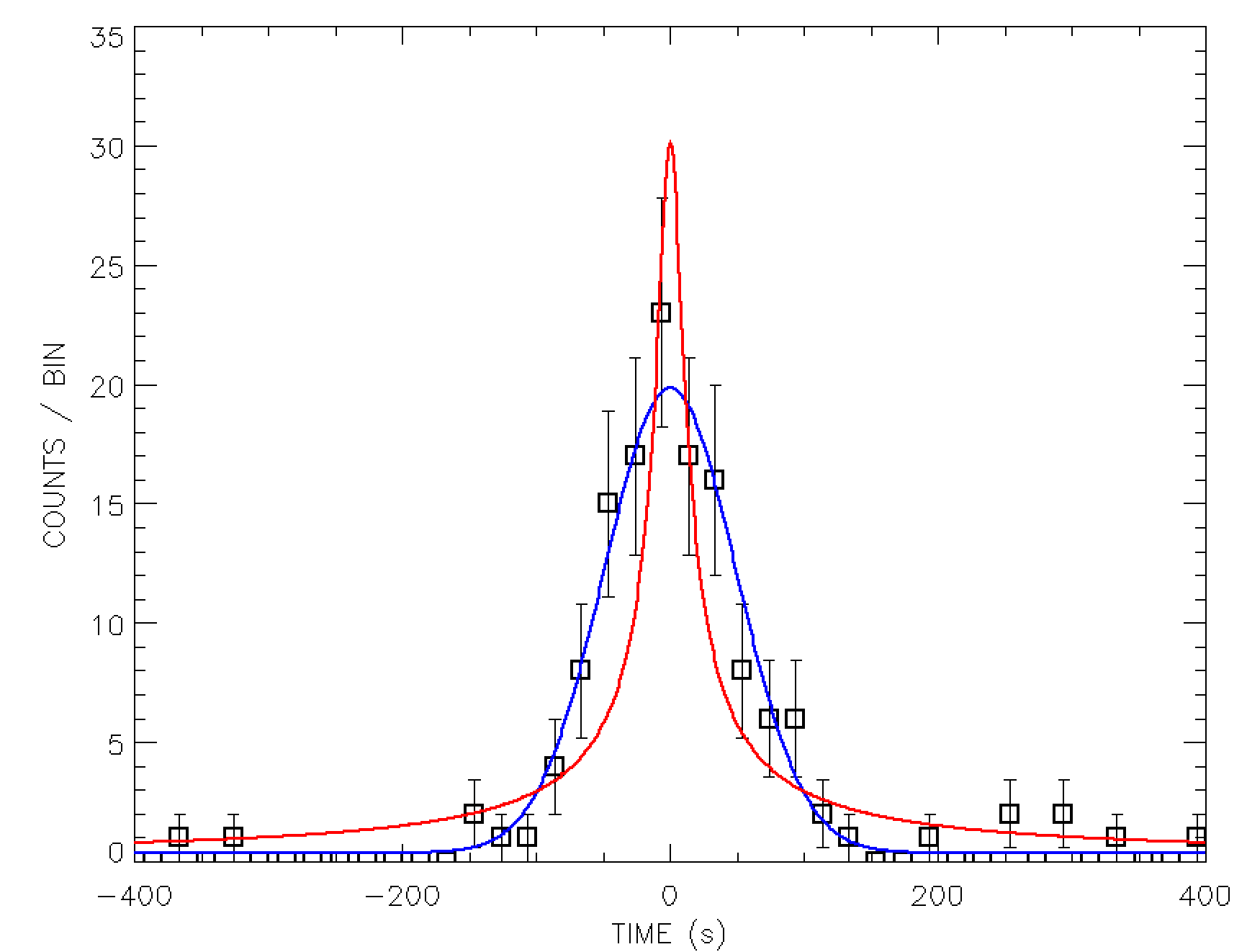}
\caption{Light curve (in 20 s bins) of the X-ray flare from \src\ observed with \xmm\ on 2005 September 21. The blue and red lines represent ML fits with a Gaussian and a gravitational microlensing model, respectively (see text for details). 
\label{fig:lc}}    
\end{figure}

\begin{figure}
	\includegraphics[width=\hsize]{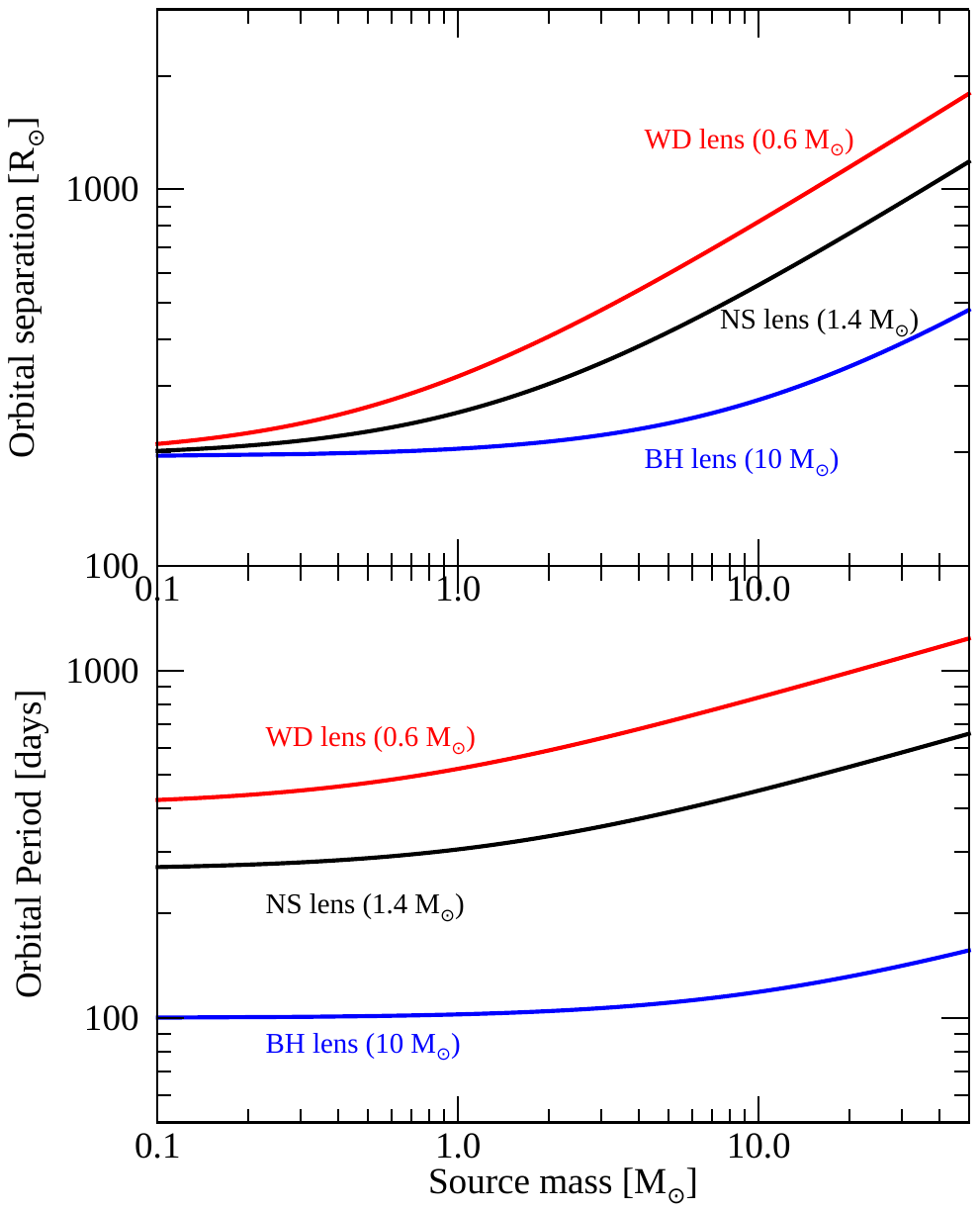}
\caption{Orbital separation (top panel) and orbital period (bottom panel) as a function of the mass of the lensed object for three representative hypotheses on the nature of the compact companion: a white dwarf of 0.6 $M_\odot$ (red), a neutron star of 1.4 $M_\odot$ (black), and a black hole of 10 $M_\odot$ (blue). \label{fig:bin}}    
\end{figure}

\subsection{Position of \src}
\label{xmm_pos}

To assess the exact position of the flaring source \src\ we reanalyzed the \xmm\ data using a maximum likelihood (ML) method, as described in \citet{rigoselli18}, to properly account for the presence of sources N and S. We based our analysis on the data of the EPIC-MOS cameras in the 0.5--5 keV energy range, as the EPIC-pn data indicated a gap between CCDs just in correspondence with source N.

We first performed a ML fit using the whole observation to simultaneously derive the positions of sources \src, N, and S in instrumental coordinates. These were then registered using the \cxo\ positions of sources N and S  given in Table~\ref{tab:src}. We finally applied the ML fit to the image corresponding only to the time interval of the flare to obtain the position of \src: R.A. = 271.5379, Dec = $-27.7653$ (1$\sigma$ error of 0.75" on R.A. and 0.77" on Dec). Applying the same analysis procedure to different energy ranges and to the EPIC-pn data, we obtained consistent results.

\subsection{Other X-ray sources in \ngc}

In addition to sources A, B, and C, there are six other X-ray sources within the extension of \ngc\ detected above a $3\sigma$-threshold by \texttt{wavdetect} (see Fig.~\ref{fig:im_gc}). Their corrected coordinates, given in Table~\ref{tab:src}, were determined using the same procedure described above. The two brightest sources, labeled N and S, were also detected in the \xmm\ observation that recorded the flare from \src. We extracted their spectra from the longest \cxo\ observation and fitted them with \texttt{Xspec} \citep{arnaud96}, modeling their emission with an absorbed power law with free $N_\textup{H}$. The resulting best-fit parameters are reported in Table~\ref{tab:src}. We note that for these two sources -- with enough statistics to perform a spectral fit -- the \nh\ value was left free to vary. The resulting values are compatible with the Galactic absorption in the direction of \ngc\,, thus corroborating our decision to fix the column density to this value for the other sources as well. The fluxes for the other sources reported in Table \ref{tab:src}  were obtained by fixing their photon index to 1.5. Sources D and F, which possess the brightest optical counterparts, are likely foreground stars.
Sources N and D show significant variability between the three most recent \cxo\ observations. Source N varies by a factor of $\approx3$ between the first two and the most recent one. Source D shows an increase in its flux by a factor of $\approx10$, and, given its optical counterpart, is likely associated with a stellar flare.\\
   
\section{Discussion} 
\label{sec:discussion}

Figure \ref{fig:HST} shows the positions of sources~A, B, and C plotted on the optical HST image of the central part of the GC \ngc. Here, we indicate the error region of  \src\ (Sect.~\ref{xmm_pos}) and the position of the source detected in 2008 with \cxo, as previously reported in \citet{mereghetti18}. The improved location accuracy of our new \cxo\ data reduces the number of possible optical counterparts of source A, which, considering its position and brightness, is the most likely origin of the 2005 flare seen by \xmm. Its flux measured with \cxo\ is lower than the quiescent level seen with \xmm, implying long-term variability by a factor $\sim7\pm2$. However, the quiescent flux of \src\ was probably contaminated by the contributions of sources B and C, which could not be resolved with the angular resolution of \xmm. Taking this possibility into account, no strong claims for long-term variability can be made. Regardless of which \cxo\ source is the quiescent counterpart of \src, the properties of the 2005 flare defy easy classification. Based on the properties described so far, we discuss different scenarios for the emission of \src.

\rev{\subsection{Gravitational lensing}}

\rev{The} remarkably symmetric time profile \rev{of the flare}, with virtually identical rise and decay times (see Fig.~\ref{fig:lc}), is reminiscent of variability caused by gravitational lensing.  The chances that this was caused by an unbound compact object passing in front of the X-ray source are extremely small. Therefore, considering the large number of binary systems with compact objects in globular clusters,  we explored the more likely possibility of binary self-lensing.

Using a ML method to properly account for the small number of counts in each bin, we fitted the flare light curve (Fig.~\ref{fig:lc})  with the magnification model expected in the approximation of a point-like source and lens. The best-fit parameters derived with the microlensing model are $\tau_e=R_E/v=906$ s and $u_0$=0.012, where $R_E$ is the Einstein radius, $v$ is the lens transverse velocity with respect to the Earth-source direction, and $u_0$ is the impact parameter (see, e.g., \citealt{paczynski86,gould92}). Even if the flare light curve is better described by a Gaussian, we can use the best-fit parameters of the microlensing model combined with Kepler's law to constrain the parameters of the binary system. 
This is shown in Fig.~\ref{fig:bin}, where the size and orbital period are plotted for three representative compact objects as a function of the companion star mass. Wide systems with orbital periods longer than 200 d for a neutron star lens (or more than 100 d if the lens is a black hole of 10 $M_\odot$) are required.

Beyond the goodness-of-fit issue lies the challenge of creating a physical model within which binary self-lensing can produce an event with the observed properties. Would such a wide binary survive in the dense central region of a globular cluster? Only a subset of those consistent with the light curve fit are ``hard'' binaries, with orbital velocities higher than ambient stellar velocities. Hard binaries are less likely to be torn apart by interactions with other stellar systems in the dense environment near the cluster center, but developing consistent physical models for such systems is even more challenging. Furthermore, the high required system inclination, extremely low probability of observing the event at the right orbital phase, and possible hints of spectral variability between the flare and quiescence lead us to discard the binary self-lensing hypothesis. 

\rev{\subsection{IMBH flare}}

The symmetry in the rise and fall times of the X-ray flare observed by \xmm\ is unusual, as most X-ray flares observed in a galaxy have a fast-rise, exponential decay (FRED) shape. This profile is indicative of rapid energy injection during the rise in luminosity, followed by slower thermodynamic cooling during the decline in luminosity. Very few flare mechanisms can lead to a symmetric profile other than the lensing model described above. However, symmetric X-ray flares are sometimes seen from the supermassive black hole Sgr A$^{\star}$ \citep{porquet08,neilsen13,haggard19}. This similarity is sufficient to motivate a further discussion of this model. At first glance, it appears possible that the X-ray flare seen in NGC~6540 could be a scaled-down version of the symmetric flares seen in Sgr A$^{\star}$, i.e., from an IMBH in NGC~6540. The energy output from representative flares of Sgr A$^{\star}$ \citep[$0.7-3.3 \times 10^{39}$ erg; ][]{haggard19} is about 700-3000 times greater than the total energy output of the \src\ flare, which would imply an IMBH mass of a few thousand ($1.3-6\times10^3$) solar masses. In addition, the fitted power-law X-ray indices of the quiescent and flare emission spectra of Sgr A$^{\star}$ ($\Gamma=3.0 \pm 0.2$ and $2.0 \pm 0.2$, respectively; \citealt{haggard19}), are remarkably similar to the quiescent and flare emission power-law values of J1806 ($\Gamma=2.5 \pm 0.5$ and $1.7 \pm 0.2$, respectively \citealt{mereghetti18}). 

If confirmed, \src\ would represent the first IMBH in a GC for which the quiescent emission is directly observed \citep[see, e.g., ][for a review]{greene20}. Assuming an X-ray bolometric correction of 10, we infer that the accretion rate of this source is $\lambda_\textup{Edd}=L_\textup{bol}/L_\textup{Edd}\approx2\times10^{-9}(M_\textup{IMBH}/10^3\textup{M}_\odot)^{-1}$. This value is far below the Bondi rate for typical values of gas densities in GCs and accretion efficiencies, but similar to the values inferred for $\omega$~Cen \citep[$\log\lambda_\textup{Edd}<-12$][]{haberle24} \revv{and Sgr~A$^\star$ \citep[$\log\lambda_\textup{Edd}\sim10^{-7}-10^{-9}$][]{quataert99}}. This might suggest a low accretion efficiency, a lower gas density, a combination of these two factors, or an intrinsically different accretion state with respect to ``standard'' accreting BHs.

However, there are \rev{several} major inconsistencies with an IMBH interpretation for the J1806 flare. The most striking feature is that J1806 lies 7$^{\prime\prime}$ (0.14 pc) from the dynamical center of \ngc. Although the cluster is rather heavily obscured and somewhat ill-defined in shape, it seems unlikely that the error in the position center could be as large as 7$^{\prime\prime}$, and dynamical friction would very quickly cause a 5,000 M$_{\odot}$ BH to sink to the dynamical center of the cluster. Also, if the emission mechanism causing the $\sim$1 hour flares seen in Sgr A$^{\star}$ (interpreted as magnetic recombination near the innermost stable circular orbit of the black hole) scales \rev{linearly} with mass, it would imply a flare duration of only $\sim$10 seconds for a 5,000 M$_{\odot}$ black hole. This is far shorter than the observed $\sim$5 minute duration of J1806's flare. Although a flare emanating from material at a larger radius ($\sim$100 gravitational radii) in the disk of a 3,000 M$_{\odot}$ black hole could have an orbital period of a few minutes, it would cool radiatively (rather than through adiabatic expansion and rapid synchrotron radiation, as in Sgr A$^\star$) and therefore exhibit a FRED profile. In addition, we lack independent evidence (e.g., from stellar dynamics and surface brightness profiles) that \ngc\ hosts such a massive IMBH. Furthermore, the presence of an IMBH with a mass of a few $10^3$~M$_\odot$ in a GC is itself somewhat problematic. While different groups have suggested -- via N-body simulations -- that dense star clusters could host IMBHs with masses of $100-500$~M$_\odot$ \citep{arcasedda21,dicarlo21,rizzuto22,gonzales22}, achieving heavier masses might prove challenging through mergers or accretion. This is despite the claim that the IMBH hosted in $\omega$ Cen could have a mass $\gtrsim8000$~M$_\odot$ \citep{haberle24}. A lighter IMBH would imply an even stronger tension with the timescales of the flares. Thus, an explanation involving a Sgr A$^{\star}$-like flare from an IMBH in NGC~6540 has significant issues.

Finally, the fact that source A has two neighbors within 4$^{\prime\prime}$ is surprising, given that there are only nine sources within 1$^{\prime}$ of the cluster center. Two of the nine sources are likely foreground or background sources for the 1$^{\prime}$ region considered (\citealt{lehmer12}), leaving $\sim$7 sources associated with the cluster itself. We performed 100,000 simulations by randomly placing seven sources in a 1$^{\prime}$ region using a King profile with a core radius of 1.8$^{\prime\prime}$ (appropriate for NGC~6540; \citealt{harris96}). We find that only $\sim$1\% of the simulations contained three sources located at a distance of at least three core radii from the cluster center. Although this could simply be fortuitous, the grouping of two sources (one of which shows significant long-term variability) in such proximity to a seemingly unique X-ray flaring source is intriguing.\\

\rev{\subsection{Background AGN flare}}

The possibility that \src\ is associated with a background active galactic nucleus (AGN) flare is extremely unlikely for several reasons. First and foremost, considering the typical $\log N-\log S$ of AGNs \citep[e.g.,][]{georgakakis08,mateos08}, we expect $2\times10^{-3}$ background AGNs at the same flux level to fall, by chance alignment, within 7" from the center of \ngc. Additionally, among the various classes of AGN-related X-ray flares, quasiperiodic eruptions (QPEs; e.g., \citealt{miniutti19}) are arguably the phenomenon most morphologically similar to the flaring behavior seen in \src, in terms of variability and pulse shape. However, QPEs are characterized by a considerably softer X-ray spectrum than the flare observed by \xmm, making this association even more unlikely. Taken together, these arguments allow us to confidently exclude a background AGN origin for the flare.\\

\section{Conclusions}
 \label{sec:conclusion}

In this work, we presented a reanalysis of the X-ray emission from \ngc, exploiting both archival \xmm\ data and $\approx65$~ks of new \cxo\ observations. The superior angular resolution of \cxo\ is critical here: it resolves what appeared as a single \xmm\ source into three distinct sources (A, B, and C), with A being the most likely counterpart of the 2005 flare, and demonstrates that the quiescent \xmm\ flux was likely the result of blended emission from the newly resolved sources. The origin of the remarkably symmetric X-ray flare, however, remains uncertain. A binary self-lensing event is disfavored by the unphysical orbital parameters, the low a priori probability it requires, and the traces of spectral variability in the flare with respect to the quiescent emission. We considered an IMBH interpretation; however, this is inconsistent with the source's offset from the dynamical center of the cluster and the flare's expected timescale. Finally, we excluded the possibility of \src\ being associated with a background AGN flare, based on its X-ray spectral appearance and location. Despite this lack of a clear physical explanation, our work represents a significant step forward in characterizing the X-ray source population of \ngc, disentangling blended emission that was previously unresolvable, and placing meaningful constraints on the nature of one of the most unusual X-ray flares observed in a Galactic globular cluster. This ongoing puzzle calls for dedicated theoretical modeling efforts to identify viable physical mechanisms capable of reproducing the observed phenomenology.

\begin{acknowledgements}
We thank the anonymous referee for precious comments that greatly improved the quality of the manuscript. We acknowledge financial support from INAF Fundamental Research Grants through the  “Toward Neutron Stars Unification”  Program (PI S.Mereghetti). JAI acknowledges support from \cxo\ grant GO3-24063X issued by the Chandra X-ray Center, which is operated by the Smithsonian Astrophysical Observatory for and on behalf of the National Aeronautics and Space Administration under contract NAS8-03060. AS acknowledges logistical support from IUSS Pavia. This research has made use of data obtained from the Chandra Data Archive and the Chandra Source Catalog, both provided by the Chandra X-ray Center (CXC). This paper employs a list of Chandra data sets, obtained by the Chandra X-ray Observatory, available at~\url{https://doi.org/10.25574/cdc.628}.

\end{acknowledgements}

\bibliographystyle{aa} 
\bibliography{biblio} 

\end{document}